\documentclass[prl,twocolumn,showpacs,floatfix,amsmath,amssymb,superscriptaddress]{revtex4-1}
\usepackage{amsfonts}
\usepackage{stmaryrd}
\usepackage{bbm}
\usepackage{mathrsfs}
\usepackage{tipa}
\usepackage{amssymb}
\usepackage{txfonts}
\usepackage{graphicx}
\usepackage{dcolumn}
\usepackage{epstopdf}
\usepackage[colorlinks,linkcolor=blue,urlcolor=blue,citecolor=blue]{hyperref}
\usepackage{multirow}
\usepackage{subfigure}
\usepackage{url}
\usepackage{upgreek}

\begin{document}
\newcommand*{\cm}{cm$^{-1}$\,}

\title{Light-induced new collective modes in La$_{1.905}$Ba$_{0.095}$CuO$_4$ superconductor}

\author{S. J. Zhang$^\star$}
\affiliation{International Center for Quantum Materials, School of Physics, Peking University, Beijing 100871, China}

\author{Z. X. Wang$^\star$}
\affiliation{International Center for Quantum Materials, School of Physics, Peking University, Beijing 100871, China}

\author{L. Y. Shi}
\affiliation{International Center for Quantum Materials, School of Physics, Peking University, Beijing 100871, China}

\author{T. Lin}
\affiliation{International Center for Quantum Materials, School of Physics, Peking University, Beijing 100871, China}

\author{M. Y. Zhang}
\affiliation{International Center for Quantum Materials, School of Physics, Peking University, Beijing 100871, China}

\author{G. D. Gu}
\affiliation{Condensed Matter Physics and Materials Science
Department, Brookhaven National Lab, Upton, New York 11973, USA}

\author{T. Dong}
\email{taodong@pku.edu.cn}
\affiliation{International Center for Quantum Materials, School of Physics, Peking University, Beijing 100871, China}

\author{N. L. Wang}
\email{nlwang@pku.edu.cn}
\affiliation{International Center for Quantum Materials, School of Physics, Peking University, Beijing 100871, China}
\affiliation{Collaborative Innovation Center of Quantum Matter, Beijing, China}

\begin{abstract} We report near and mid-infrared pump c-axis terahertz probe measurement on a superconducting single crystal La$_{1.905}$Ba$_{0.095}$CuO$_4$ with T$_c$=32 K. The measurement reveals that the pump-induced change occurs predominantly at the Josephson plasma edge position below T$_c$. Upon excited by the strong near-infrared pulses, the superconducting state is severely disturbed and incoherent quasiparticle excitations develop in frequency regime above the static plasma edge. However, within very short time delay ($\sim$1.5 ps) we observe the reappearance of a very sharp Josephson plasma edge at frequency lower than the static Josephson plasma edge and the emergence of a new light-induced Josephson mode at higher energy. The results imply that the light can induce new Josephson couplings with different coupling strengths. Similar but weaker effect is observed for the mid-infrared pump. No pump induced effect is detected above T$_c$.
\end{abstract}


\maketitle

Recent development of ultrashort laser pulses has been proven as a powerful tool for light control of different orders in complex electronic materials. Paradigmatic examples include induction of lattice distortions in manganites\cite{Rini2007b}, transient generation of spin-density-wave order in the normal state of BaFe$_2$As$_2$\cite{Kim2012}, melting\cite{Porer2014a} and switching\cite{Stojchevska2014} of charge-density wave (CDW) orders in transition metal dichalcogenides, manipulation of the order parameters and detection of Higgs/amplitude modes in superconductor/CDW compounds\cite{Matsunaga2013,Chen2017}, light-induced electron localization in a quantum Hall system\cite{Arikawa2017}, etc. Among various novel phenomena, the light-induced superconductivity in cuprates is perhaps the most intriguing and exciting observation. The effect was first observed in a stripe-ordered cuprate at 10 K in the normal state \cite{Fausti189}, then in underdoped YBa$_2$Cu$_3$O$_{6.5}$ at temperature even above 300 K \cite{Kaiser2014,hu2014optically}. In those measurements, after excited by  mid-infrared (MIR) 15 $\upmu$m ($\sim$ 80 meV), the Josephson plasma edges formed by superfluid carriers were detected in a time-domain terahertz (THz) measurement. More recently, it was found that the transient superconductivity could be induced and enhanced by near-infrared (NIR) pump at 800 nm in La$_{1.885}$Ba$_{0.115}$CuO$_4$, which is close to the stripe-ordered phase \cite{Nicoletti2014,PhysRevB.91.174502}.

Here we report near and mid-infrared pump THz probe measurement on a superconducting single crystal La$_{1.905}$Ba$_{0.095}$CuO$_4$ with T$_c$=32 K. The sample locates at the composition where bulk superconducting transition temperature (T$_c$), charge stripe order temperature (T$_{co}$), spin stripe order temperature T$_{so}$, and structural phase transition temperature from low-temperature orthorhombic (LTO) \emph{P}4$_2$/\emph{ncm} to low-temperature less
orthorhombic (LTLO) \emph{Pccn} are all close to each other in the phase diagram \cite{PhysRevB.83.104506}. We find that, for both near- and mid-infrared pumps, the pump-induced change occurs predominantly at the Josephson plasma edge position below T$_c$. The superconducting state is severely disturbed or suppressed by the strong NIR excitations at the moment when the pump induced reflectance change of THz peak reaches maximum, as reflected by the significant suppression of the Josephson plasma edge. Meanwhile, incoherent quasiparticle excitations develop at energy higher than the static plasma edge. However, within very short time delay $\sim$1.5 ps, we observe the reappearance of a very sharp Josephson plasma edge at energy lower than the Josephson plasma edge in the static case and a new light-induced Josephson mode at higher energy. The results suggest the establishment of new Josephson couplings from superfluid carriers with different coupling strengths.  Although the pump-induced change evolves in a trend towards its equilibrium state, the decay is rather slow and sizeable effect is still seen at time delay of 50 ps after the pump excitations.

Single crystals of La$_{1.905}$Ba$_{0.095}$CuO$_4$ were grown using the traveling-solvent floating-zone method\cite{Homes2012}. Infrared spectra in the equilibrium state were measured with Bruker 113/v and 80v Fourier transform infrared spectrometers (FTIR) at different temperatures. Near- to mid-infrared pump--THz probe spectroscopy system in reflection geometry was constructed to measure static and pump-induced time-domain THz electric field \cite{Zhang2017}. A transverse electric field (TE) configuration, i.e. the THz electric field perpendicular to the incident plane and parallel to the c-axis of crystal, is employed for the measurement, as displayed in Fig. \ref{Fig:static} (a). Detailed experimental setup and measurement technique are presented in supplemental file \cite{SI}.

\begin{figure}[htbp]
  \centering
\includegraphics[width=8.5cm]{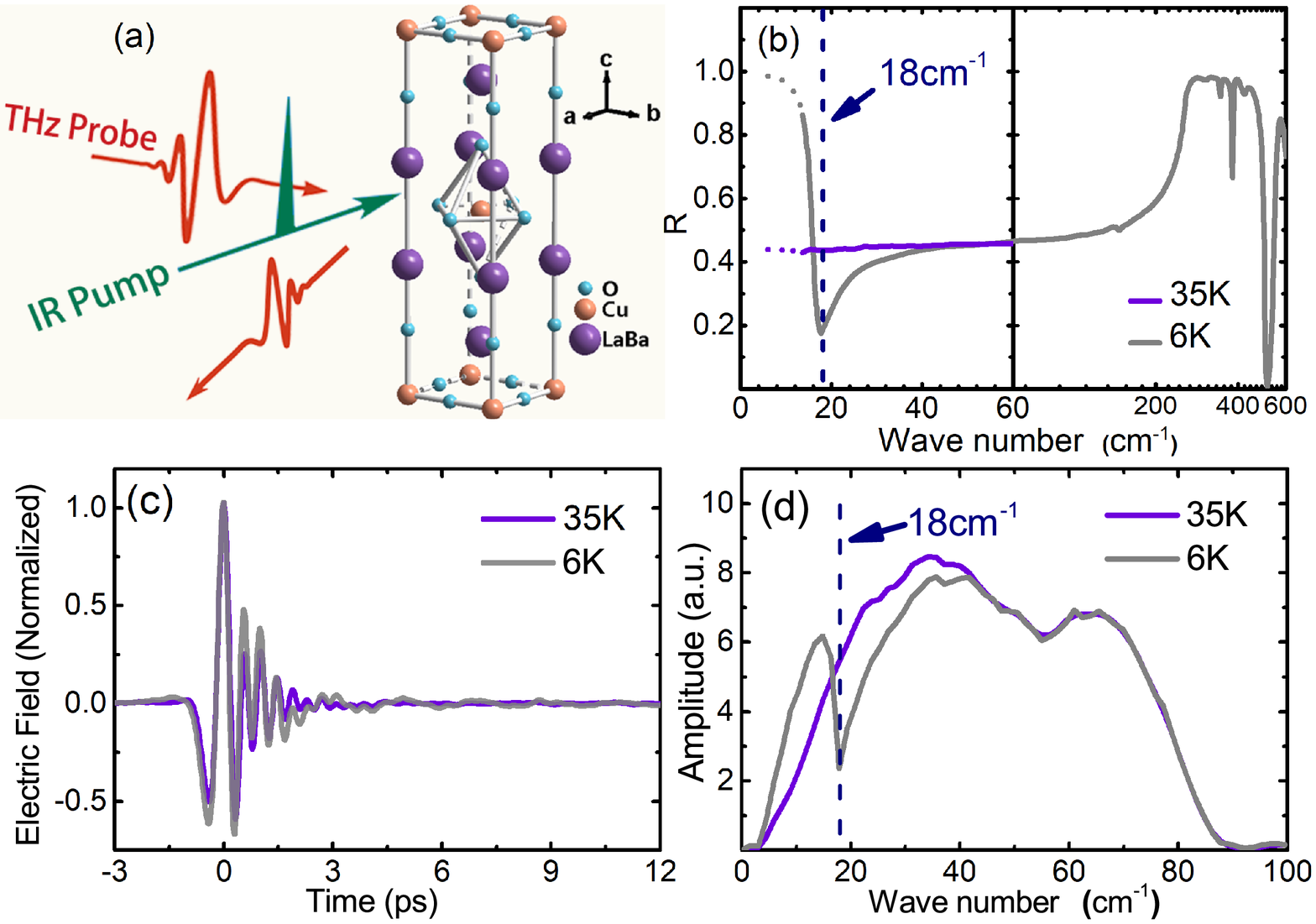}\\
  \caption{ Crystal structure and equilibrium optical spectra. (a) Both the pump and THz probe electric fields are polarized to the c-axis of the sample. The THz spectrum is measured in a transverse electric field configuration.  (b) The far infrared reflectance spectra measured by FTIR at two selective temperatures 35 K and 6 K above and below T$_c$, respectively. The dot curves are extended by THz measurement. (c) and (d) Static THz spectra in time-domain $E(t)$ and their Fourier transformed spectra in frequency domain $E(\omega)$ at 35 K and 6 K. Dash line indicates the plasma edge dip position. }\label{Fig:static}
\end{figure}

Figure \ref{Fig:static} (b) shows the far-infrared (FIR) reflectivity spectra with the electric field of light E//c-axis at two selective temperatures measured by FTIR. The spectra show a number of phonon modes in FIR region. Above T$_c$, the reflectivity values are low and relatively featureless below 100 cm$^{-1}$ (or 3 THz), indicating insulating response. A very sharp Josephson plasma edge is clearly observed near 18 cm$^{-1}$ ($\sim$0.54 THz) at 6 K below T$_c$. The data are in good agreement with earlier measurement by Homes \cite{Homes2012}. The reflectance extension to lower energy (dot curves in Fig. \ref{Fig:static} (b)) is achieved by the time-domain THz measurement. Figure \ref{Fig:static} (c) and (d) shows static THz spectra in time-domain $E(t)$ and their Fourier transformed spectra in frequency domain $E(\omega)$ at two selective temperatures 35 K and 6 K above and below T$_c$, respectively. The sharp dip near 18 \cm at 6 K corresponds to the Josephson plasma edge observed in the FTIR measurement. The presence of the Josephson plasma edge is a hallmark of superconductivity in layered cuprates.

\begin{figure}[htbp]
  \centering
\includegraphics[width=8.5cm]{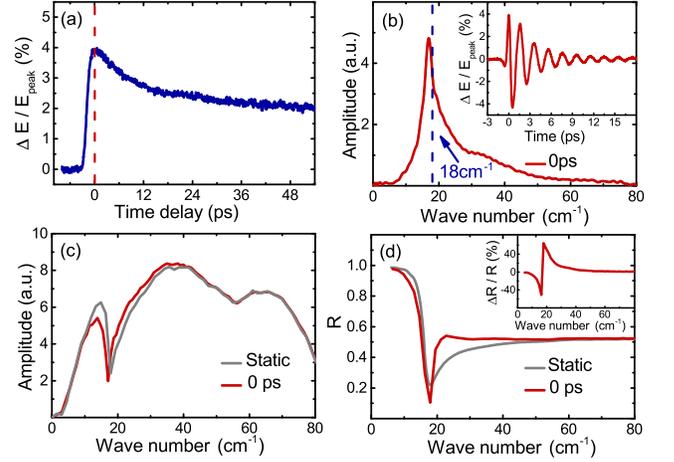}\\
  \caption{Pump-induced changes. (a) The relative electric field change at the THz peak position as a function of time delay after excitation at 6 K. (b) Inset: the pump induced relative change $\Delta E(t)/E_{peak}$ in time domain at $\tau$=0 ps. Main panel: the Fourier transformed spectrum of $\Delta E(t)$ in frequency domain. (c) Frequency domain THz spectrum before and after excitation at $\tau$=0 ps. (d) The reflectivity spectra before and after excitation at $\tau = 0$ ps. Inset shows the ratio of the reflectivity change relative to the static values. The penetration depth mismatch is not considered here.}\label{Fig:DeltaE}
\end{figure}

We now show the time resolved terahertz measurement with NIR pump at 1.28 $\mu m$ ($\sim$ 1 eV). Figure \ref{Fig:DeltaE} (a) displays the decay of the relative change of THz electric field peak, $\Delta E/E_{peak}$, after excitation by a fluence of 3 mJ/cm$^2$. The decay process lasts for long time and sizable signal is still present at time delay of 50 ps after excitation. At each time delay after excitation, the pump-induced change in the reflected THz electric field $\Delta E(t)$ can be acquired directly by modulating the pump pulse.
Figure \ref{Fig:DeltaE} (b) shows the pump induced relative change $\Delta E(t)/E_{peak}$ at 6 K below T$_c$ at maximum pump-probe signal position being defined as time delay $\tau$=0 ps ( inset). The pump-induced THz signal shows clearly oscillations in the time domain, which gives a pronounced peak slightly below 18 \cm in the frequency domain $\Delta E(\omega,\tau)$ after Frourier transformation, as shown in the main panel of Fig. \ref{Fig:DeltaE} (b). This pronounced peak in $\Delta E(\omega, \tau)$ suggests that the pump-induced change occurs predominantly near the static Josephson plasma edge position. A comparison of frequency domain THz spectrum after excitation, $E(\omega, \tau)$+$\Delta E(\omega, \tau)$, with that of static THz spectrum, $E(\omega, \tau)$, is shown in Fig. \ref{Fig:DeltaE} (c). Obviously, the electric field is reduced below the static plasma edge and enhanced just above the plasma edge for $\tau$=0 ps.

The complex reflection coefficient of the photo-excited sample, $\tilde{r}'(\omega,\tau)$, can be determined
from the normalized pump-induced changes to the electric field  $\Delta\tilde{E}(\omega,\tau)/\tilde{E}(\omega)$ using the relation
 $\Delta\tilde{E}(\omega,\tau)/\tilde{E}(\omega)$=[$\tilde{r}'(\omega,\tau)$-$\tilde{r}(\omega)$]/$\tilde{r}(\omega)$, where $\tilde{E}(\omega)$ and $\Delta\tilde{E}(\omega,\tau)$ are obtained from the Fourier transformation of measured $E(t)$ and $\Delta E(t,\tau)$,  the static reflection coefficient $\tilde{r}(\omega)$ is evaluated from the equilibrium
optical reflective index obtained from Kramers-Kronig transformation of optical reflectance measured by FTIR. From $\tilde{r}'(\omega,\tau)$, we can calculate all the optical constants. Figure \ref{Fig:DeltaE} (d) shows the reflectivity spectrum after excitation at $\tau = 0$ ps. The static reflectivity is also plotted for comparison. We find that
the reflectivity is suppressed below the static plasma edge and enhanced just above the plasma edge, and merged into the static values roughly above 40 \cm. Those features can be more clearly seen from the plot of the ratio of reflectivity change over the static values, as shown in the inset of Fig. \ref{Fig:DeltaE} (d).

\begin{figure*}[htbp]
  \centering
  \includegraphics[width=15cm]{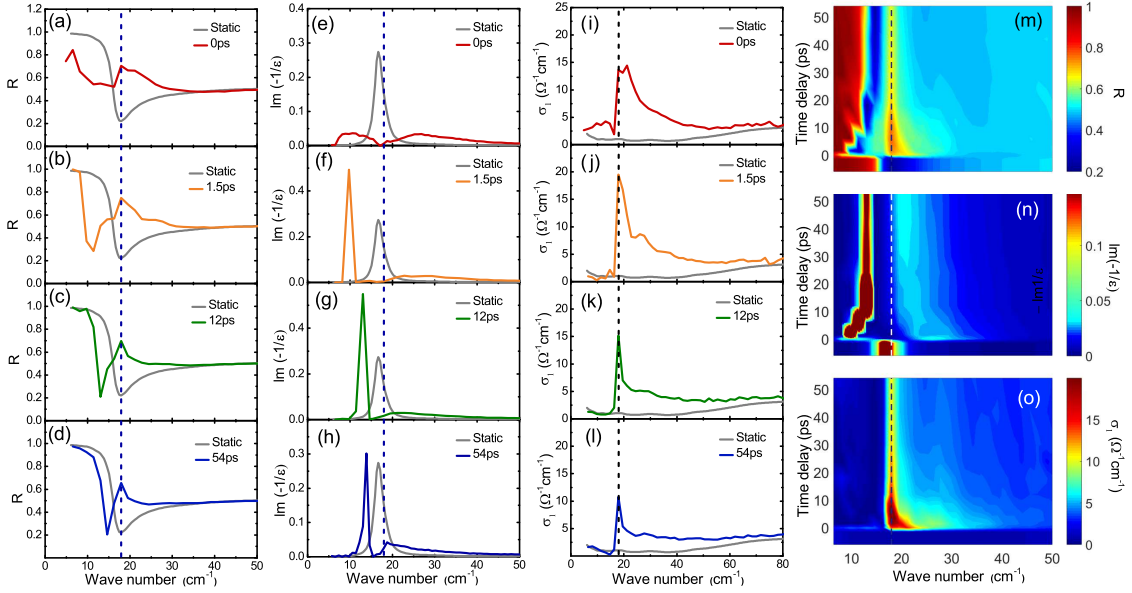}\\
  \caption{Calculated optical constants from a multilayer model at different time delays after 1.28 $\upmu$m excitations. (a)-(d) show the spectral changes at several representative time delays. It can be seen that the Josephson plasma edge at 18\cm is suppressed to lower energy and  a new plasma edge appears at higher energy after 1.28 $\upmu$m excitations. (e)-(h) show the corresponding energy loss function spectra, in which those two longitudinal Josephson plasmon modes are much easier to identify. (i)-(l) show the real part of conductivity spectra after excitations. The peak features observed in the real part of conductivity spectra indicate the appearance of a transverse mode. (m), (n) and (o) are intensity plot of the spectral evolution of R($\omega$), Im(-1/$\varepsilon(\omega)$) and $\sigma_1(\omega)$ as a function of time delay. }\label{Fig:15mw}
\end{figure*}

It deserves to emphasize that there exists a significant difference in penetration depths between NIR pump (at 1.28 $\upmu$m) and the THz probe pulses (below 2.5 THz). To derive the true pump-induced THz spectral change, the penetration depth mismatch should be taken into account. The data shown above are the raw experimental measurement results of static and pump-induced THz spectra together with some calculated quantities at a specific time delay without considering such mismatch. In the following, we present calculated optical constants from a multilayer model by assuming the pumped volume as a stack of thin layers, in which the  refractive index, $\widetilde{n}$, evolves as a function along the  direction of propagation. The incident angle of 30$^0$ is also taken into account in the calculations. Detailed procedure is described in the supplementary file \cite{SI}.

Figure \ref{Fig:15mw} shows the calculated reflectivity, energy loss function and real part of conductivity spectra at different time delays after 1.28 $\upmu$m excitations. We first examine the spectral changes at several representative time delays. At the maximum pump-induced signal position $\tau$=0 ps ( Fig. \ref{Fig:15mw} (a)), the reflectivity below the edge is severely suppressed to low values (roughly below 0.8), though it still displays an edge-like shape. By contrast, the values near the dip position are strongly enhanced and a new heavily overdamped plasma edge appears at higher energy before merging into the static spectrum. The results suggest that the superconductivity is strongly disturbed and suppressed. However, after a very short time delay, e.g. at $\tau$=1.5 ps  ( Fig. \ref{Fig:15mw} (b)), a very sharp plasma edge appears at much lower frequency near 10 \cm, then a flat reflectance forms followed by a new broad enhancement with an overdamped edge-like feature at higher energy. With increasing the time delay, e.g. $\tau$=12 and 54 ps  ( Fig. \ref{Fig:15mw} (c) and (d)), the sharp plasma edge at lower energy scale shifts to higher energy scale, meanwhile the broad enhancement at high energy weakens and a sharp edge feature appears near 18 \cm. The observations indicate that the superconductivity is strongly suppressed by the initial photo excitations, but within short time delay new Josephson plasmons with different coupling strengths reestablish. The two edges, which have been well formed after short time delay of photo-excitations, represent two longitudinal Josephson plasmon modes. They can be identified more clearly as peaks in the energy loss function spectra as presented in Fig. \ref{Fig:15mw} (e)-(h), where the locations of the peaks indicate the energy scales of the mode and their widths reflect the damping rates of the collective excitations. We remark that the splitting of the pump induced plasma mode was observed in YBa$_2$Cu$_3$O$_{6.45}$  with 15 $\mu m$ excitation \cite{Hunt2016}, however the signal level was very small and visible only in the differential spectra between the pumped and static spectra. Formation of two extremely strong and sharp light-induced longitudinal plasmon modes as presented here was not seen before.

Presence of two longitudinal Josephson plasmon modes would lead to the formation of a
transverse Josephson plasmon between the two longitudinal modes, which can be regarded as an out-of-phase oscillation
of the two individual components \cite{VanderMarel1996} and has been observed in many cuprate systems\cite{Shibata1998,Grueninger1999,Timusk2003,Tajima2012}. The transverse mode shows up directly in conductivity spectrum. Indeed, a peak feature is clearly observed in real part of conductivity spectra, as presented in  Fig. \ref{Fig:15mw} (i)-(l). The peak becomes very sharp and locates at the frequency just slightly lower than the energy of the higher plasmon mode as indicated in the energy loss function spectra. Besides the narrow peak, a broad enhancement is observed in conductivity above the static Josephson plasmon energy, which is associated with the strongly enhanced reflectivity near and above the static plasma edge energy. This feature is prominent at $\tau$=0 ps but becomes weaker at longer time delays. This broad feature suggests development of incoherent quasiparticle excitations with binding energy above the static plasma edge energy. More detailed spectral evolutions of R($\omega$), Im(-1/$\varepsilon(\omega$)) and $\sigma_1(\omega)$ as a function of time delay are presented in the intensity plot in Fig. \ref{Fig:15mw} (m), (n) and (o), respectively.

\begin{figure}[htbp]
  \centering
\includegraphics[width=8.5cm]{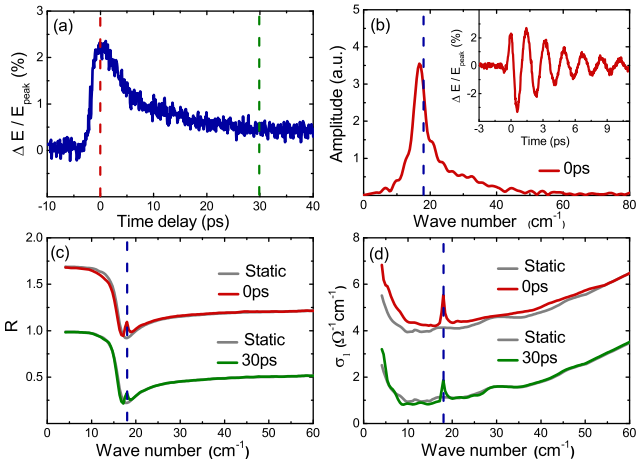}\\
  \caption{ Pump-induced changes by 15 $\upmu$m excitations at 6 K. (a) The relative electric field change at the THz peak position as a function of time delay after excitation. (b) The pump induced relative change of $\Delta E(t)/E_{peak}$ in time domain (inset) and the Frourier transformed $\Delta E(\omega,\tau)$ in the frequency domain (main panel) at $\tau$=0 ps. (c) and (d) Calculated reflectivity and real part of conductivity from a multilayer model at two time delays $\tau$=0 ps and 30 ps and their comparisons with static spectra.}\label{Fig:pump15mum}
\end{figure}

We performed similar measurement with MIR excitations at 15 $\upmu$m. Figure \ref{Fig:pump15mum} (a) shows the decay of the relative change of THz electric field at peak position $\Delta E/E_{peak}$ after excitation at a fluence of 0.5 mJ/cm$^2$. The decay process is similar to the excitation at 1.28 $\upmu$m. The pump induced relative change of $\Delta E(t)/E_{peak}$ in the time domain and the Fourier transformed $\Delta E(\omega,\tau)$ in the frequency domain at T=6 K and $\tau$=0 ps are shown in the inset and main panel of Figure \ref{Fig:pump15mum} (b), respectively. The shapes are almost the same as that of 1.28 $\upmu$m excitations though the signal level is smaller due to weak pump fluence. Figure \ref{Fig:pump15mum} (c) and (d) show the calculated reflectivity and real part of conductivity spectra from the multilayer model at two selective time delays $\tau$=0 ps and 30 ps and their comparisons with static spectra, respectively. We find that, although the pump-induced spectra are in principle similar to that excited by 1.28 $\upmu$m shown above, the effect is much smaller. Only a very small peak appears near the static plasma edge position. The high frequency side edge of the small peak  can be assigned to the second longitudinal plasmon mode. It leads to a weak feature in energy loss function (not shown) and a peak in conductivity spectrum (i.e. transverse plasmon mode). Additionally, the broad enhancement feature is not visible, only very small difference between static and pumped spectra can be seen at $\tau$=0 ps. The very small effect in calculated optical constants is attributed to smaller penetration depth mismatch between 15 $\upmu$m and THz probe beam besides the weaker pump fluence.

We also performed the pump induced THz measurement at different temperatures for both 1.28 $\upmu$m and 15 $\upmu$m excitations. The signal becomes weaker and gradually disappears with increasing temperature (see supplementary file \cite{SI}). No pump induced signal is detected above T$_c$. Our measurement results are different from earlier measurement on the same composition by Nicoletti et al. \cite{Nicoletti2014}. They observed only very small effect with 800 nm ($\sim$1.5 eV) pump at a fluence of 2 mJ/cm$^2$. A very small blue shift of the Josephson plasma edge was indicated in their calculated spectrum, suggestive of small enhancement of interlayer Josephson coupling.

We now discuss the implications of measurement results. An important pump-induced effect is the development of strong and relatively broad feature in $\sigma_1(\omega)$ at the energy scale above the static plasma edge. It reflects quasiparticle excitations. This feature is prominent with NIR pump at $\tau$=0 ps but its intensity decays with time delays. Because it is not Drude like, we assign it to the incoherent excitations. It implies that the excited quasiparticles are still confined in the ab-plane and the c-axis coherence can not be established due to the insulating block layers. The binding energy has the energy scale of Josephson coupling strength. Nevertheless, at longer time delay, the reduced spectral weight should become superconducting condensate centered at zero frequency.

The observation of two longitudinal Josephson plasma modes and a transverse plasmon is the most prominent observation in this work. It suggests the development of two inequivalent Josephson couplings along the c-axis. Its origin is not clear at present. It is known that the application of a moderate external magnetic field within CuO$_2$ plane can have a remarkable
effect on the c-axis Josephson plasma mode in underdoped YBa$_2$Cu$_3$O$_{7-\delta}$ \cite{Kojima2002,LaForge2007}. The applied magnetic field can cause inequivalent insulating layers with and without Josephson vortices, which in turn lead to modulation formation of Josephson couplings. As a result, two longitudinal Josephson plasmons form and their coupling generates a transverse mode which corresponding to the antiphase Josephson current oscillations between two inequivalent junctions \cite{Kojima2002,LaForge2007}. In the present measurement, the electric field of pump pulse at the fluence of 3 mJ/cm$^2$  reaches about 6 MV/cm at 1.28 $\upmu$m pumping. Correspondingly, the magnetic field is about 2 T. The magnetic field at 15 $\upmu$m pumping is about 0.3 T. The field may be strong enough to generate the Josephson vortices in the CuO$_2$ plane. However, it is not clear how the ultrafast pulse can generate spacial modulation along the c-axis. Additionally, the 35 femtosecond pulse might be too short to form stable vortices.

Another possibility is related to the coherent atom oscillations caused by the ultrafast pumping. For example, if the intense pump pulse can drive the out-of-plane apical oxygens to vibrate along the c-axis, it can lengthen and shorten copper-apical oxygen bonds alternately among copper oxide planes. Therefore, two different Josephson coupling strengths can develop. This would naturally explains the formation of two longitudinal Josephson plasmons and a transverse mode. Nevertheless, it is not clear whether or not such coherent apical oxygen oscillations could form in the compound, and no relevant information is reported yet in literature. Under such circumstances, the intense pumping at mid-infrared 15 $\upmu$m is expected to be more efficient than at 1.28 $\upmu$m, since its energy is closer to oxygen vibration energy. However, an even weaker effect is indicated by the 15 $\upmu$m excitations.

One may also think that the compound locates close to the charge stripe order phase in the phase diagram, the dynamical stripe fluctuation may exist in the sample. Since the charge stripes, whose directions change by 90$^o$ in the neighboring copper oxide planes, can lead to spatial modulation of the Josephson coupling strength between the superconducting charge stripes \cite{Gorshunov2011,Voronkov2012}, the  observation of two longitudinal plasma edges may be related to the dynamical stripe orders. We think that this possibility is low since the compound exhibits single and very sharp Josephson plasma edge without pumping. The intense pumping is unlikely to induce formation of charge stripe order. Apparently, further studies are necessary to elucidate the true origin.

To summarize, we performed near and mid-infrared pump c-axis THz probe measurement on a superconducting single crystal La$_{1.905}$Ba$_{0.095}$CuO$_4$ with T$_c$=32 K. The measurement reveals that the pump-induced change occurs mainly at the Josephson plasma edge position below T$_c$. The superconducting state is severely disturbed by the strong near-infrared excitations and incoherent quasiparticle excitations develop in frequency regime above the static plasma edge. Most prominently, we observe the reappearance of a very sharp Josephson plasma edge at frequency lower than the static Josephson plasma edge and a new light-induced Josephson mode at higher energy after short time delay. The results imply that light can induce new Josephson couplings with different coupling strengths. Similar but weak effect is observed for the mid-infrared pump.

\begin{center}
\small{\textbf{ACKNOWLEDGMENTS}}
\end{center}
$^\star$These authors contributed equally to this work. We acknowledge very useful discussions with Shin-ichi Uchida, Setsuko Tajima and John Tranquada. This work was supported by the National Science Foundation of China (No. 11327806, GZ1123) and the National Key Research and Development Program of China (No.2016YFA0300902, 2017YFA0302904).

\bibliographystyle{apsrev4-1}
  \bibliography{Terahertz}

\appendix

\section{Supplementary}

\section{S1 Experimental Apparatus and Measurement Methods }

All time-domain terahertz (THz) spectroscopy measurements presented in this paper were performed on a near-to mid-infrared (MIR) pump -- THz probe spectroscopy system in reflection geometry.

In this spectroscopy system, a two-output optical parametric amplifier sharing with the same white light continuum is used for pump pulse generation. The two outputs both containing signal and idler beams ranging from 1.2 to 1.6 $\mu$m and 1.6 to 2.6 $\mu$m at 35 fs duration and 1 kHz repetition rate, respectively, can be used directly as the near-infrared (NIR) pump. To obtain the MIR pump, two signal beams which have been tuned to perpendicular polarizations are used for difference frequency generation collinearly on a 1-mm-thick z-cut GaSe crystal. MIR pulses with tunable polarization ranging from 3 to 16 $\mu$m can be generated. The THz probe pulses are generated from pulses of 800 nm light whose polarization can be tuned by a $\lambda/2$-plate using a 1-mm-thick (110) ZnTe crystal without focalization. The THz beam was focused on the sample by a 30${}^{\circ}$ off-axis parabolic mirror with a spot size of 0.63 mm and detected via electro-optic sampling using 1-mm-thick ZnTe. Detailed experimental setup and measurement technique were presented elsewhere \cite{Zhang2017}.

Broadband Infrared spectra in the equilibrium state were measured with Bruker 113/v and 80v Fourier transform infrared spectrometers (FTIR) at different temperatures and the equilibrium optical constants can be calculated by Kramers--Kroenig transformation from the spectra.

 \section{S2 Multilayer Model Used for Transient Optical Constants Calculation}

\begin{figure*}[htbp]
  \centering
\includegraphics[width=15cm]{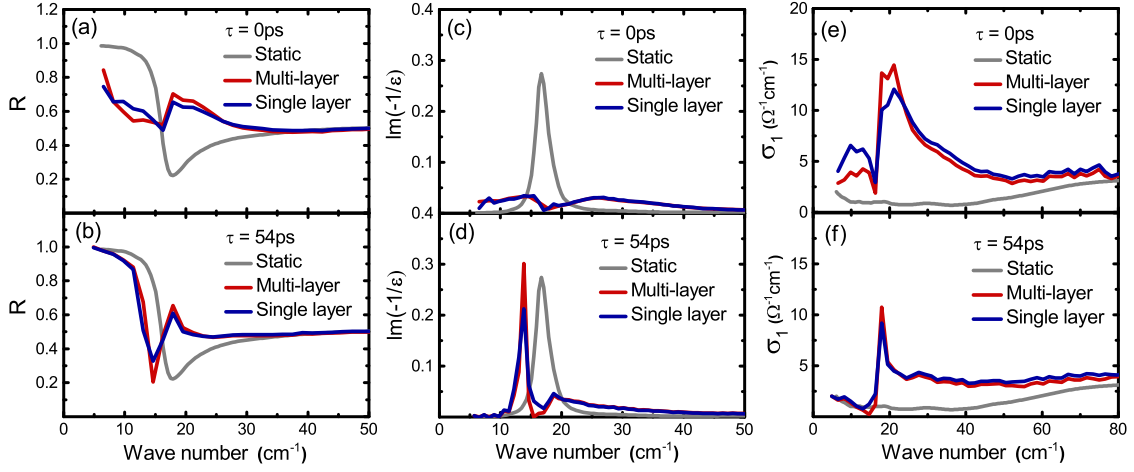}\\
  \caption{Comparison of the single layer and multilayer models. (a) and (b) show the spectral changes at $\tau$ = 0ps and 54ps calculated from the single layer and multilayer models. The suppression of the Josephson plasmon edge and the induce of a new plasma edge can be clearly seen in both models. (c) and (d) show the corresponding energy loss function spectra, in which those two longitudinal Josephson plasmon modes can be identified in both models. (e) and (f) show the real part of conductivity spectra after excitations, in which the peak features are both clear in single layer and multilayer models.}\label{Fig:modelcompare}
\end{figure*}

As we mentioned in the main text, there exists a significant difference in penetration depths between the NIR pump at 1.28 $\mu$m (or mid-infrared at 15$\mu$m) and the THz probe pulses (below 2.5 THz) due to the fact that the pump pulses have much higher energy than the probe pulses. The penetration depths mismatch must be taken into account in deriving the actual pump-induced optical constants. This can be achieved by a multilayer model or single layer model. The calculated optical constants presented in the main text of this work are obtained by the multilayer model.

For a thin homogeneous layer with a thickness of $z$, the material can be specified by a characteristic matrix $\mathbf{M}(z)$ according to the Maxwell's equations and appropriate boundary conditions. When an electromagnetic wave, whose wavelength is $\lambda_{0}$ in vacuum, is incident on a non--magnetic layer, the characteristic matrix $\mathbf{M}(z)$ can be written as \cite{born2013principles}
\begin{equation}
\mathbf{M}(z)=\begin{bmatrix} cos(k_{0}\tilde{n}zcos\theta_{0})& -\frac{i}{p}sin(k_{0}\tilde{n}zcos\theta_{0}) \\ -{i}{p}sin(k_{0}\tilde{n}zcos\theta_{0})&cos(k_{0}\tilde{n}zcos\theta_{0}) \end{bmatrix}\quad
\end{equation}
 where $\theta_{0}$ is the angle of incidence and $k_{0} = 2\pi/\lambda_{0}$. For transverse electric field configuration, $p=\tilde{n}cos\theta_{0}$.
 
 In the multilayer model, it is assumed that many homogeneous thin layers with evolving $\tilde{n}$ stack together along the direction of propagation, which is denoted as $z$ direction. The characteristic matrix of the total probe-interrogating region can be written as a product of the matrices for each layer,
\[\mathbf{M}(z_{n})=\mathbf{M_{1}}(z_{1})\mathbf{M_{2}}(z_{2}-z_{1}) \cdots \mathbf{M_{n}}(z_{n}-z_{n-1}).\]
Detailed formulas can be found in \cite{born2013principles}.

In the multilayer model, it is assumed that $\tilde{n}(z) = \tilde{n}_{0} + \Delta \tilde{n}\cdot e^{-z/l_{p}}$, where the latter term is the pump-induced change of refractive index which follows the same decay law as the light intensity. It exponentially decays to $\Delta\tilde{n}/e$ when $z$ goes to the penetration depth of pump pulses $l_{p}$. Hence the characteristic matrix of this total probed region can be written as
\[\mathbf{M}(z_{n}) = \begin{bmatrix} 1& -ik_{0}l_{p} \\
-ik_{0}l_{p}(\tilde{n}_{0}^{2}+2(1-e^{-L/l_{p}})\tilde{n}_{0}\Delta \tilde{n} + \frac{1-e^{-2 L/l_{p}}}{2}\Delta \tilde{n}^{2}-sin^{2}\theta_{0})
&1 \end{bmatrix}\quad,\]
where $L$ is the penetration depth of THz probe pulses whose wavelength is $\lambda_{0}$ in vacuum.

We denote the vacuum as medium 1, the pumped region of the material as medium 2, and the unpumped region as medium 3. $\tilde{r}_{13}$ is the equilibrium complex reflection coefficient which can be acquired by FRIT measurements and the Kramers--Kroenig transformation, and $\tilde{r}_{12}$ is the pump-induced transient complex reflection coefficient considering the penetration depth mismatch which is waiting for the following calculation. The reflection coefficient acquired by time-domain THz spectroscopy measurements $\tilde{r}$ can be expressed by the elements of	the	$\mathbf{M}(z_{n}) $, $m_{ij}$, as below \cite{born2013principles}
\begin{equation}
\tilde{r}=\frac{(m_{11}+m_{12}p_{3})p_{1}-(m_{21}+m_{22}p_{3})}{(m_{11}+m_{12}p_{3})p_{1}+(m_{21}+m_{22}p_{3})}
\end{equation}
where $p_{1} = cos\theta_{0}$ and $p_{3}=\frac{1-\tilde{r}_{13}}{1+\tilde{r}_{13}}cos\theta_{0}$.
Hence, $\Delta \tilde{n}$ can be solved as an unknown of a quadratic equation shown below
\[\begin{split}\frac{1-e^{-2L/l_{p}}}{2}ik_{0}l_{p}\cdot\Delta \tilde{n}^{2}+2(1-e^{-L/l_{p}})ik_{0}l_{p}n_{0}\cdot\Delta \tilde{n}\\
+(1-ik_{0}l_{p}p_{3})p_{1}\frac{1-\tilde{r}}{1+\tilde{r}}+(ik_{0}l_{p}n_{0}^{2}-ik_{0}l_{p}sin^{2}\theta _{0}-p_{3})=0.\end{split}\]
There may exist two roots for the quadratic equation according to the quadratic formula. The way to pick a reasonable solution is to maintain the real part of $\tilde{n}'= \tilde{n}_{0} + \Delta \tilde{n}$ positive, for the calculated results should keep in line with the definition of physical quantities. The pump-induced transient optical properties shown in Fig. 3 and Fig. 4 can be obtained by $\tilde{n}' $.

We can also derive a single layer model from the above characteristic matrix $\mathbf{M}(z)$ by assuming that the total reflection coefficient is a combined contribution from the photoexcited top layer with a constant complex refractive index and the non-excited bottom bulk. The reflection coefficient can be obtained by inserting the matrix elements in Equ. (1) into Equ. (2). It has the form
\[\tilde{r}=\frac{\tilde{r}_{12}+\frac{\tilde{r}_{13}-\tilde{r}_{12}}{1-\tilde{r}_{12}\tilde{r}_{13}}exp(2i\delta)}{1+\tilde{r}_{12}\frac{\tilde{r}_{13}-\tilde{r}_{12}}{1-\tilde{r}_{12}\tilde{r}_{13}}exp(2i\delta)},\]
where $\delta=2\pi l_p \tilde{n'}/\lambda_0$. The transient optical properties of La$_{1.905}$Ba$_{0.095}$CuO$_4$ calculated from the single layer and multilayer models are shown in Fig. \ref{Fig:modelcompare}. It can be seen that the pump-induced changes of optical properties calculated from both models are qualitatively similar. Only minor difference exists in magnitude.

\begin{figure}[htbp]
  \centering
\includegraphics[width=8cm]{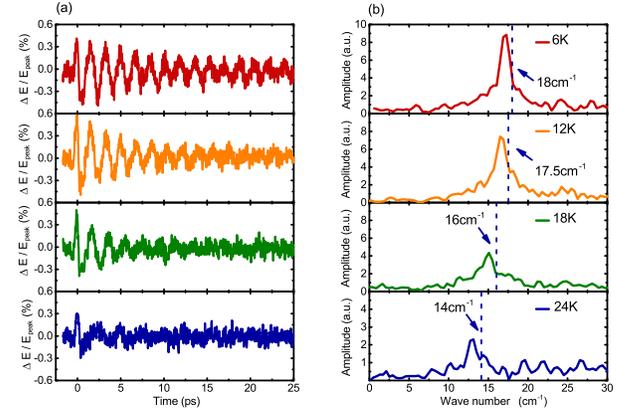}\\
  \caption{Transient changes of reflective electric field at different temperature.  (a) shows pump--induced changes in time domain after 15 $\mu$m excitations at different temperatures below $T_{c}$. (b) shows the corresponding Fourier transformed spectrum, in which a peak slightly below the redshifting static Josephson plasma can be seen.}\label{Fig:Temp}
\end{figure}

 \section{S3 Temperature dependence measurements}
Figure \ref{Fig:Temp} (a) displays the pump induced relative changes $\Delta E(t)/E_{peak}$ at $\tau$=0 ps after 15 $\mu$m excitations at a fluence of 0.2mJ/cm$^2$. The pump-induced THz signals with clearly oscillations in time domain can also be identified at higher temperatures below $T_{c}$. The corresponding  Fourier transformed spectrum are shown in Figure \ref{Fig:Temp} (b). The navy dash lines were denoted as the static Josephson plasma edge dip positions at different temperatures characterized by THz	time-domain spectroscopy, which had a redshift as the temperature warming up. A pronounced peak slightly below the redshift edge position can be seen in the frequency domain, which was similar with the case at 6K.

Similar experiment results were found in a temperature dependence measurements with the $1.28 \mu m$ excitation cases. No pump-induced THz signals can be seen above $T_{c}$ even when the fluence is over 3mJ/cm$^2$. The absence of pump-induced signal may be related to the peculiar overlap of the bulk superconducting transition temperature T$_c$, charge stripe order temperature T$_{co}$, spin stripe order temperature T$_{so}$, and structural phase transition temperature of  La$_{1.905}$Ba$_{0.095}$CuO$_4$.


\end{document}